\documentclass[%
reprint,
showpacs,showkeys
 amsmath,amssymb,
 aps,
]{revtex4-1}

\usepackage{epsfig}
\usepackage{graphicx}
\usepackage{dcolumn}
\usepackage{bm}

\usepackage{hyperref}

\begin{document}
\def\ca{$\sim$}
\def\b209{$^{209}$Bi }
\def\b209n{$^{209}$Bi}
\preprint{APS/123-QED}

\title{First measurement of the partial widths of $^{209}$Bi decay to the ground and to the first excited states.\\}

\author{J.~W.~Beeman$^{1}$}
\author{M.~Biassoni$^{2,~3}$}
\author{C.~Brofferio$^{2,~3}$}
\author{C.~Bucci$^{4}$}
\author{S.~Capelli$^{2,~3}$}
\author{L.~Cardani$^{5}$}
\author{M.~Carrettoni$^{2,~3}$}
\author{M.~Clemenza$^{2,~3}$}
\author{O.~Cremonesi$^{3}$}
\author{E.~Ferri$^{2,~3}$}
\author{A.~Giachero$^{3}$}
\author{L.~Gironi$^{2,~3}$}
\author{P.~Gorla$^{4,~6}$}
\author{C.~Gotti$^{3,~7}$}
\author{A.~Nucciotti$^{2,~3}$}
\author{C.~Maiano$^{2,~3}$}
\author{L.~Pattavina$^{3}$}
\author{M.~Pavan$^{2,~3}$}
\author{G.~Pessina$^{3}$}
\author{S.~Pirro$^{3,~4}$}
\author{E.~Previtali$^{2,~3}$}
\author{M.~Sisti$^{2,~3}$}
\author{L.~Zanotti$^{2,~3}$}

\affiliation{$^{1}$Lawrence Berkeley National Laboratory, Berkeley, California 94720, USA}
\affiliation{$^{2}$Dipartimento di Fisica, Universit\`a di Milano-Bicocca, Milano I-20126 - Italy}
\affiliation{$^{3}$INFN - Sezione di Milano Bicocca, Milano I-20126 - Italy}
\affiliation{$^{4}$INFN - Laboratori Nazionali del Gran Sasso, Assergi (L'Aquila) I-67010 - Italy}
\affiliation{$^{5}$Dipartimento di Fisica, Sapienza Universit\`a di Roma, and INFN- Sezione di Roma, Roma I-00185, Italy}
\affiliation{$^{6}$INFN - Sezione di Roma II, Roma I-00133 - Italy}
\affiliation{$^{7}$Dipartimento di Elettronica e TLC, Universit\`a di Firenze, Via S. Marta
3, Firenze I-50125 - Italy}

\date{\today}

\begin{abstract}
$^{209}$Bi alpha decay to the  ground and to the first excited state have been contemporary observed for the first time with a large BGO scintillating bolometer. The half-life of $^{209}$Bi is determined to be $\tau_{1/2}$=(2.01$\pm$0.08)$\cdot$10$^{19}$~years while the branching ratio for the ground-state to ground-state transition is (98.8$\pm$0.3)\%.

\end{abstract}

\maketitle

\section{INTRODUCTION\\}
$^{209}$Bi is the only naturally abundant isotope of bismuth. Its $\alpha$-decay to $^{205}$Tl was predicted on the basis of the $^{209}$Bi mass excess, but escaped direct observation until 2003 when it was observed with a small BGO scintillating bolometer, Ref.~\cite{nature}. In that work -- due to the small mass of the detector -- only the partial width for the ground state decay was measured, resulting in a half-life of (1.9$\pm$0.2)$\times$10$^{19}$~yr. The decay was in that case ascribed to $^{209}$Bi on the basis of the energy measured for the emitted $\alpha$ line, interferences from other $\alpha$-decaying isotopes -- like $^{190}$Pt and $^{186}$Os -- were excluded making reasonable assumptions on the purity of the crystal.  
However, the decay can proceed either as a direct transition to the $^{205}$Tl ground state (GS-GS transition) or as a transition to the first excited level of $^{205}$Tl (GS-ES transition), see Fig.~\ref{fig:209Bidecay}. The contemporary observation of both the lines would provide -- as observed also in Ref.~\cite{nature} -- the real conclusive test on the identification of $^{209}$Bi decay.

In this paper we report the first experimental evidence for $^{209}$Bi decay to the 204~keV excited level of $^{205}$Tl, the measurement of the branching ratio for the two transitions and finally the evaluation of $^{209}$Bi half-life.

\section{EXPERIMENTAL TECHNIQUE}
The detector used for this study is a BGO scintillating bolometer. This is realized instrumenting a Bi$_4$Ge$_3$O$_{12}$ crystal with a temperature sensor and coupling it to a Light Detector (LD). Both the BGO crystal and the LD operate as bolometers, at a temperature of few tens of mK. The working principle is quite simple: the energy released by an ionizing particle that traverses the BGO crystal is converted both into scintillation light and into heat. The former gives rise to a light pulse that is recorded by the LD, the latter produces a temporary temperature increase of the BGO crystal that is recorded by its temperature sensor. The fraction of the total deposited energy spent in scintillation depends on the nature of the particle and is called Light Yield. $\beta$'s and $\gamma$'s have the same LY, which is typically different from that of $\alpha$'s or neutrons; in this way heat and light signals can be used to disentangle particle identity. At room temperature BGO crystals produce about (8-10)$\cdot$10$^3$ scintillation photons for a 1 MeV electron, corresponding to a LY$_{\gamma}$ of the order of 20-26~keV/MeV. The scintillation yield increases when cooling the crystal~\footnote{see for example Saint Gobain data-sheets}.

\subsection{Detectors and experimental set-up}

The design of the single BGO bolometer and of the LD follows that of the other macro-bolometers operated in the past by our groups (Ref.~\cite{cadmio,seleniuro}).
The main bolometer is a 5x5x5~cm$^2$ BGO crystal (with a mass of 889.09~g) purchased from SICCAS (China) with optically grade surfaces. 
The crystal is secured inside a copper holder and is surrounded on 5 faces by a light reflector (3M VM2002). Attached to the crystal is an NTD Ge thermistor (Ref.~\cite{NTD}) that acts as thermometer: the temporary temperature increase of the crystal -- following particle interaction -- produces a voltage pulse with an amplitude proportional to the deposited energy. 
The LD is a bolometer built with the same technique and optimized in order to be able to detect the small energy carried by the scintillation photons. The LD consists in a high purity Ge wafer, 36~mm diameter and 0.3~mm in thickness. On the side facing the scintillating crystal the surface of the wafer is `darkened' with a deposition of a 600~\AA\ layer of SiO$_2$ to improve light absorption. An NTD Ge thermistor for the temperature signal read-out is glued on the other side of the LD, to which a $^{55}$Fe source for energy calibration is faced.

The BGO and the LD bolometers (namely the heat and light channels) are provided with two completely independent read-out chains. The voltage signal produced on the NTD thermistors are amplified and filtered by the front-end electronics and fed into the ADC. When the trigger fires, the entire signal waveform is sampled, digitized and saved to disk.
Pulse height and pulse-shape parameters are computed by the off-line analysis for each acquired event. Signals deformed by excess noise or by pile-up are rejected by the analysis since they can produce a broadening or a deformation of the peaks, spoiling the energy resolution. The true rate is then reconstructed measuring the trigger efficiency and the probability that a pulse-shape cut rejects a true particle event. The former is measured on pulser generated events, the latter is obtained by comparing the event rate in a background peak before and after pulse-shape cuts (the ratio between the two rates is the pulse-shape cut efficiency). 
 
\begin{figure}[t]
\begin{center}
\includegraphics[width=0.8\linewidth]{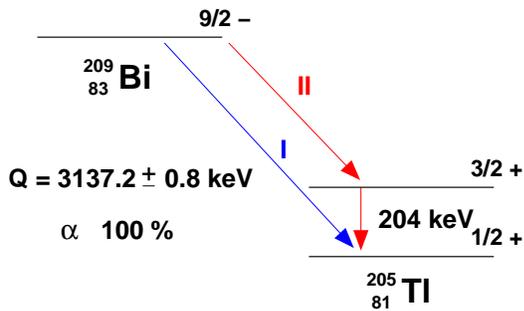}
\end{center}
\caption{Decay scheme of $^{209}$Bi, data from Ref.~\cite{nucldata}. The GS-GS transition (I) occurs with the emission of a monochromatic $\alpha$ line of 3077~keV. The GS-ES transion (II) occurs with the emission of an $\alpha$ particle of 2876~keV and is followed by the emission of a 204~keV gamma ray.}
\label{fig:209Bidecay}
\end{figure}

\subsection{Energy calibration}
The energy content of the heat and light signals are evaluated using calibration data.

The light signal is calibrated using $^{55}$Fe X-rays, assuming a linear dependence of the pulse height on energy. This is an approximation since in general the relationship between energy and pulse height is not linear (the exponential dependence of the thermistor resistance on temperature being the main source of non-linearity).
The calibration returns the energy content of the recorded scintillation signal. It should be noted that the energy is not corrected for the light collection efficiency, therefore it does not provide an absolute determination of the amount of scintillation light produced by the BGO crystal.

The heat signal is calibrated using the more intense $\gamma$ lines visible in the BGO heat spectrum. They cover a range from 0.5 to 2.6~MeV and are produced by radioactive contamination of the crystal or of the cryogenic set-up. Above and below the calibration is obtained by extrapolation. It is important to remark that in calibrating the heat signal the nominal $\gamma$ line energy is assumed for each full energy peak observed in the heat spectrum. This means that the calibration -- for $\beta$/$\gamma$ events -- returns the total amount of energy released by the interacting particle in the crystal (although this energy is only partially converted into heat). On the contrary for $\alpha$ events it overestimates the true total energy, since the LY of $\alpha$ particles (LY$_{\alpha}$) is different (lower) than for $\beta$/$\gamma$ (LY$_{\gamma}$) (for a detailed discussion of this issue see~\cite{cadmio}). To 
emphasize this fact, the heat axis units are indicated as keV$_{\beta/\gamma}$.

\section{RESULTS}
The detector was operated in the cryogenic facility installed in Hall C of Laboratori Nazionali del Gran Sasso (L'Aquila, Italy). The BGO crystal cooled very slowly (while the cryostat -- as usual -- reached the base temperature in few days), a behavior that can be ascribed to some excess heat capacity that decouples at low temperatures. 

We collected 374.6 hours of background measurement in semi-stable conditions: the crystal was at 30~mK, still cooling during the measurement period improving its signal height by $\sim$ 30\%. Off-line corrections -- using the technique described in Ref.~\cite{impulsatore} -- were applied to account for the thermal drift observed during the measurement.

\begin{figure}[t]
\begin{center}
\includegraphics[width=1\linewidth]{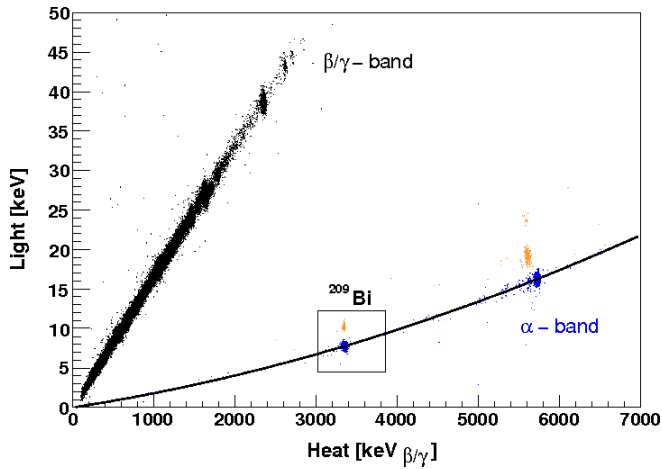}
\end{center}
\caption{Light vs.heat scatter plot corresponding to 374.6~hours of background (no external sources) measurement. The heat axis is calibrated with gamma lines, as described in the text;  the subscript $\beta$/$\gamma$ on the heat  `keV' units indicates that the calibration gives the correct energy only for $\beta$/$\gamma$ particles. Colors are used to highlight the $\alpha$ band. Pure $\alpha$-decays lie on a curve that is here fitted with a degree two polynomial. Mixed $\alpha$+$\gamma$ events lie above this curve, they are highlight in a lighter color. In the inset $^{209}$Bi decay events.}
\label{fig:BIGscatt}
\end{figure}

\begin{figure}[t]
\begin{center}
\includegraphics[width=1\linewidth]{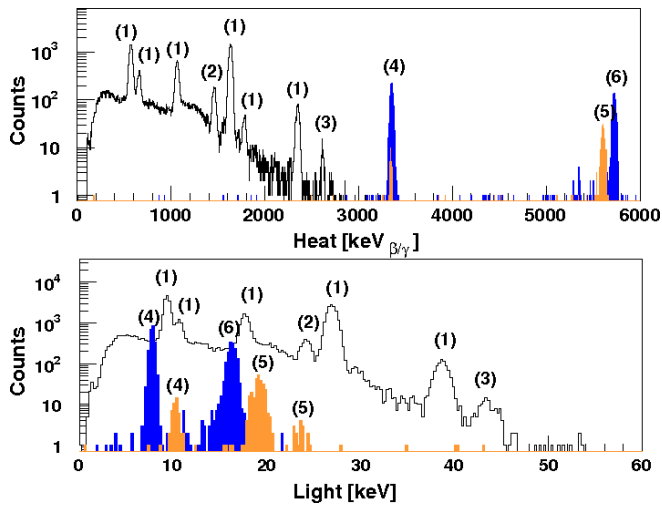}
\end{center}
\caption{Heat (top) and light (bottom) spectra of events belonging to the $\beta$/$\gamma$ band (black histrogram) and $\alpha$ band (colored filled histograms). Numbers are used to identify the main lines: (1) is for $\gamma$ lines of $^{207}$Bi, (2) for those of $^{40}$K and (3) for $^{232}$Th, (4) indicates the $\alpha$ lines of $^{209}$Bi, (5) those of $^{210}$Bi, (6) that of $^{210}$Po.}
\label{fig:BIGspectra}
\end{figure}

The light vs. heat scatter plot corresponding to this statistics is shown in Fig.~\ref{fig:BIGscatt}. The $\alpha$ and $\gamma$ regions appear to be clearly separated, their light and heat spectra are reported in Fig.~\ref{fig:BIGspectra}.

\subsection{Gamma region}
Several very intense lines are visible in the $\gamma$ spectra (and were used -- as discussed above -- for the heat spectra energy calibration). They are due to the internal contamination of the crystal in $^{207}$Bi and the background $^{40}$K and $^{232}$Th lines (usually observed in all measurements and ascribed to detector+cryostat contamination). $^{207}$Bi is a common contaminant already observed in BGO crystals which is produced by cosmic ray protons interaction on $^{206}$Pb, Ref.~\cite{207Bi,207BiB}.
The average FWHM energy resolution on the heat channel -- as measured on the more prominent $\gamma$ lines -- is (37.5$\pm$0.5)~keV, with no evident dependence on energy.

The energy resolution of the LD, measured on the $^{55}$Fe line, is 0.5~keV. The FWHM of the scintillation peaks produced by the $^{207}$Bi, $^{40}$K and $^{232}$Th $\gamma$ photons interacting in the BGO ranges from 0.8~keV to 1.5~keV. The LY$_{\gamma}$ has no evident dependence on energy, its average value is $\overline L\overline Y_{\gamma}$=(16.61$\pm$0.02)~keV/MeV. The overbar here indicates that LY$_{\gamma}$ is not corrected for the light collection efficiency. 

\subsection{Alpha region}

The structure of the $\alpha$ region appears slightly more complicated than the $\beta$/$\gamma$ one. Here we can identify two different kind of events. 

\underline{Pure $\alpha$-decays} (no $\gamma$ emission) are aligned along the same curve in the scatter plot (see Fig.~\ref{fig:BIGscatt}). These events are produced by $\alpha$ particles impinging on the crystal from an external source, or by $\alpha$-decays in crystal bulk. The former have generally a continuous energy distribution (see for example Ref.~\cite{ArtRadio}) the latter are monochromatic with an energy corresponding to the Q-value of the decay (since both the energies of the emitted alpha and of the recoiling nucleus are detected). Two such lines are clearly evident in our scatter plot and are identified as due to $^{209}$Bi$\rightarrow ^{205}$Tl decay (Q=3137.2$\pm$0.8~keV, Ref.~\cite{nucldata}), and to $^{210}$Po$\rightarrow^{206}$Pb decay (Q=5407~keV) which is present in the crystal, probably as a result of $^{209}$Bi activation (see Ref.~\cite{210Pb}). Since $\alpha$ particles have a LY that is lower with respect to the $\beta$/$\gamma$ one, the two lines appear in the heat spectrum with energies higher than the nominal ones. We recall that this effect is simply due to the procedure adopted for the heat axis calibration. 
The LY$_{\alpha}$ and the corresponding quenching factors (QF=LY$_{\alpha}$/LY$_{\gamma}$) are:
\begin{displaymath}
^{209}Bi~\left\{ \begin{array}{l} 
\overline L\overline Y_{\alpha}=2.482\pm0.002~keV/MeV \\
QF_{\alpha}=0.1494\pm0.0002
\end{array} \right.
\end{displaymath}

\begin{displaymath}
^{210}Po~\left\{ \begin{array}{l} 
\overline L\overline Y_{\alpha}=3.011\pm0.003~keV/MeV \\
QF_{\alpha}=0.1813\pm0.0002
\end{array} \right.
\end{displaymath}

The increase with energy of the $\overline L\overline Y_{\alpha}$ is responsible of the curvature of the $\alpha$ band, an effect already observed by Ref.~\cite{nature,coron}. In Fig.~\ref{fig:BIGscatt} we show the result obtained fitting the $\alpha$ band with a degree 2 polynomial. 

We note that $^{209}$Bi and $^{210}$Po are internal contaminations of the crystal. In principle their LY could differ from that of a pure $\alpha$ particle because a fraction of the total energy (about 2\% ) is carried by the nuclear recoil (R). However, the difference in the LY of $\alpha$ and $\alpha$+R events is far below our sensitivity. For this reason, here and in the following, we will assume $\alpha$ and $\alpha$+R events as having the same LY.

\underline{$\alpha$-decays on the excited state} of the daughter nucleus give rise to mixed $\alpha$+$\gamma$ events. In this case, the light signal is higher than what expected for a pure $\alpha$ emission, producing therefore events that lie in between the $\alpha$ and the  $\beta/\gamma$ band. 
An example is the $^{210m}$Bi decay, a contaminant responsible of the events appearing in the  mixed $\alpha$+$\gamma$ band, just at the left of the $^{210}$Po (pure-$\alpha$) line. $^{210m}$Bi is produced in bismuth by thermal neutron interaction and it accumulates in the material due to its long half-life (see Ref.~\cite{210Pb}). The isotope $\alpha$-decays (Q=5036.4 keV) to different excited levels of the daughter isotope ($^{206}$Tl) with the contemporary emission of one or more $\gamma$ ray. The two spots (visible in Fig.~\ref{fig:BIGscatt}) are ascribed to the decays to the two lowest levels (with a sum BR of 94.5\% and energies of 265.6 or 304.9~keV) and to two higher ones (with a BR of 1.4\% and 3.9\% and energies of 634.5~keV and 649.6~keV). The probability of full containment of the emitted gammas is high, ranging from (84.2$\pm$0.4)\% for the lowest energy $\gamma$ to (54.5$\pm$0.3)\% for the highest one (the efficiencies were evaluated with a GEANT-4 simulation in which the disexcitation cascades for the four considered levels were reproduced). The measured intensities agree with the tabulated branching ratios within one standard deviation. 

\subsection{$^{209}$Bi decay}

\begin{figure}[t]
\begin{center}
\includegraphics[width=1\linewidth]{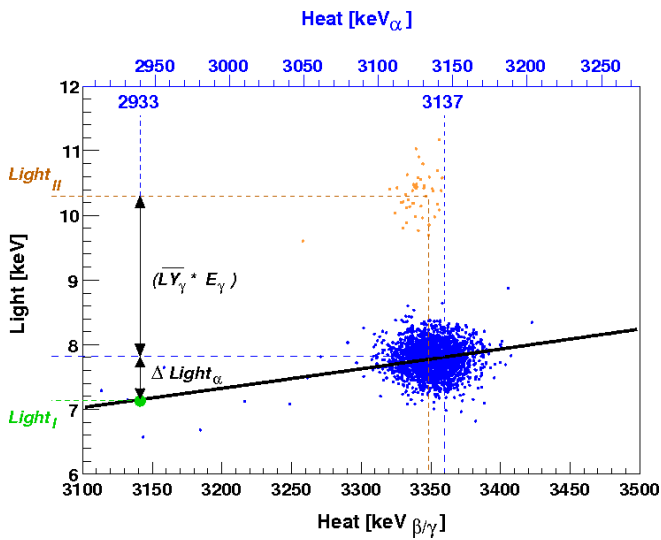}
\end{center}
\caption{Zoom of the light vs. heat scatter plot in correspondence of $^{209}$Bi events. The projections of the two spots corresponding to processes I and II along the heat axis (top panel) and the light axis (lateral panel) are shown. The horizontal axis reported on the top of the plot is obtained with a linear calibration based on the $^{209}$Bi 3137~keV pure-$\alpha$ line. The circle indicates the position that should correspond to the $\alpha$+recoil emitted in process II (2933~keV). Colors follows the convention used for figure~\ref{fig:BIGscatt}.}
\label{fig:zoom}
\end{figure}

Similar to $^{210m}$Bi is the case of our interest: $^{209}$Bi decay. It follows two different paths to the $^{205}$Tl ground state labeled I and II in Fig.~\ref{fig:209Bidecay}. 
Process I produces an $\alpha$ particle plus a recoil (Q=3137~keV); being a monochromatic pure $\alpha$-decay it should produce a line in the $\alpha$-band (the probability of fully contain both the $\alpha$ particle and the recoiling nucleus inside the crystal, $\epsilon_{\alpha}$, is obviously \ca1).
Process II produces an $\alpha$ particle plus recoil (Q=2933~keV) and a prompt $\gamma$ ray (E$_{\gamma}$=204~keV). In $\epsilon_{\gamma + \alpha}$=(92.1$\pm$0.5)\% of cases the $\gamma$ photon is fully absorbed in the BGO crystal (toghether with the $\alpha$ particle and the recoiling nucleus). In such cases we have a monochromatic $\alpha$+$\gamma$ event that should produce a line in the mixed $\alpha$+$\gamma$ band.

This is exactly what we observe in our data: two spots corresponding to roughly the same heat position (3350~keV on the $\beta$/$\gamma$ calibration scale), but a different light emission, are clearly visible (see Fig.~\ref{fig:BIGscatt}). 
Zooming the scatter plot in the $^{209}$Bi region (Fig.~\ref{fig:zoom}) we can observe that not only the light position of the two spots is different, but also the heat one. Indeed, although the emitted particles carry the same total energy, in process II a larger fraction of it is spent in the production of scintillation light. The difference in the light position is:
\begin{displaymath}
(Light_{II}-Light_{I})=\overline L \overline Y_{\gamma} \cdot E_{\gamma} - \Delta Light_{\alpha}
\end{displaymath}
where $\Delta Light_{\alpha}$ is the difference between the light signal of the 3137~keV $\alpha$+R of process I and that of the 2933~keV $\alpha$+R of process II. This relationship
can be used to estimate the energy of the emitted photon, further proving that we are observing $^{209}$Bi decay.
 
Indeed, since the non-linearity of the heat axis is very small, we can linearly calibrate it in the proximity of the $^{209}$Bi peak (in Fig.~\ref{fig:zoom} the heat axis calibrated in this way is drawn on the top). Using this calibration, we can obtain the heat position expected for an $\alpha$+R event of 2933~keV. This results (3133.8$\pm$0.3)~keV (the position is indicated with a circle in Fig.~\ref{fig:zoom}). Then, using the fit of the pure $\alpha$ band, we compute the value of $\Delta Light_{\alpha}$. Finally, we obtain for E$_{\gamma}$ a value of (192$\pm$8)~keV, validating our hypothesis.
\\
The branching ratio for the two $^{209}$Bi decays are obtained fitting the light spectrum with the sum of two gaussians having the same width and intensities: [(1-BR)$\cdot \epsilon_{\gamma + \alpha}$] for the intensity of the GS-ES transition and [BR$\cdot \epsilon_{\alpha}$] for the intensity of the GS-GS transition. The result is a BR of (98.8$\pm$0.3)\% for the GS-GS transition. The total number of events corresponding to the decay is 2199$\pm$66, with a detection efficiency of (87$\pm$2)\% (which accounts for both the trigger efficiency and the pulse-shape cuts efficiency) this yields a half-life for $^{209}$Bi nucleus of $\tau_{1/2}$=(2.01$\pm$0.08)$\cdot$10$^{19}$~years. 
Finally the GS-GS transition partial width is $\tau_{1/2}^{I}$=(2.04$\pm$0.08)$\cdot$10$^{19}$~years in good agreement with the previously reported one, Ref.~\cite{nature}.

\section{CONCLUSION} 

This work provides a compelling evidence of the observation of $^{209}$Bi decay through the contemporary observation of the ground state and excited state transitions. While confirming the half-life of the isotope already measured by Ref.~\cite{nature}, we were also able to add a new experimental information on $^{209}$Bi, namely the Branching Ratio between the ground state and the excited state transition. This study covers a seldom explored field, namely that of hindered $\alpha$-decays providing experimental inputs for a better development of the theoretical framework of nuclear models, Ref.~\cite{teorico}. In the mean time, these results proves once more the potentialities of the bolometric technique in the study of rare nuclear processes. 

\bibliography{BGO}

\end{document}